\documentclass[aps, prx,twocolumn,longbibliography,superscriptaddress,showpacs,amsmath,amssymb,floatfix]{revtex4}
\usepackage[latin9]{inputenc}
\usepackage{amssymb}
\usepackage{graphicx}
\usepackage{amsmath}
\usepackage{color}
\usepackage{mathrsfs}
\usepackage{float}
\usepackage{indentfirst}
\usepackage{mathrsfs}
\usepackage{float}
\usepackage{indentfirst}
\usepackage{textcomp}
\usepackage{comment}
\usepackage{mathtools}
\usepackage{natbib,hyperref}
\usepackage{soul}                                       
\usepackage{subfigure}

\begin{document}

\title{The Performance of VQE across a phase transition point in the $J_1$-$J_2$ model on kagome lattice}

\author{Yuheng Guo}
\affiliation{Key Laboratory of Artificial Structures and Quantum Control, School of
Physics and Astronomy, Shanghai Jiao Tong University, Shanghai 200240, China}

\author{Mingpu Qin}\thanks{qinmingpu@sjtu.edu.cn}
\affiliation{Key Laboratory of Artificial Structures and Quantum Control, School of
Physics and Astronomy, Shanghai Jiao Tong University, Shanghai 200240, China}

\begin{abstract} 
Variational quantum eigensolver (VQE) is an efficient classical-quantum hybrid method to take advantage of quantum computers in the Noisy Intermediate-Scale Quantum (NISQ) era. In this work we test the performance of VQE by studying the $J_1$-$J_2$ anti-ferromagnetic Heisenberg model on the kagome lattice, which is found to display a first order phase transition at $J_2 / J_1 \approx 0.01$. By comparing the VQE states with the exact diagonalization results, we find VQE energies agree well with the exact values in most region of parameters for the 18-site system we studied. However, near the phase transition point, VQE tends to converge to the excited states when the number of variational parameters is not large enough. For the system studied in this work, this issue can be solved by either increasing the number of parameters or by initializing the parameters with converged values for $J_2/J_1$ away from the phase transition point. Our results provide useful guidance for the practical application of VQE on real quantum computers to study strongly correlated quantum many-body systems. 
\end{abstract}

\maketitle

\section{Introduction}
The concept of quantum computing can be traced back to Feynman \cite{1982IJTP...21..467F}, who proposed to simulate physical systems with quantum computer in 1982. Many efficient quantum algorithms were developed later which are superior to the existing classical counterparts. Peter W. Shor developed the so called Shor's algorithm \cite{shor1994algorithms} in 1994. With Shor's algorithm, the factorization of a large integer number can be performed in polynomial time, while the required time for factorization in existing classical algorithms are all exponential. Shor's algorithm is one of the milestones in quantum computing, which demonstrates the power of quantum computing and poses a threat to the RSA cryptosystem. In 1996, Lov K. Grover developed a quantum searching algorithm \cite{grover1996fast} which has a square-root speedup for searching problems over any classical algorithm.

It is now widely believed that quantum computing provides an alternative to solve certain difficult problems which are beyond the capacity of classical computer. However, we need real hardware to unlock the power of efficient quantum algorithms. Progresses in building quantum computers \cite{2019Natur.574..505A,PhysRevLett.127.180501} has been made recently. But the buildup of practical large-scale quantum computer is still challenging today. It is believed that we are now in the ``Noisy Intermediate-Scale Quantum" (NISQ) era \cite{preskill2018quantum} and this era will last for a long time. Thus, developing suitable quantum algorithm for quantum computers in NISQ era is crucial nowadays. Due to the limited number of qubits in quantum computer in NISQ era, classical-quantum hybrid algorithm is one of the practical paradigms to take advantage of the power of both classical and quantum computers \cite{doi:10.7566/JPSJ.90.032001, callison2022hybrid}. 

Different types of classical-quantum hybrid algorithms were proposed in the last decade \cite{Bauer_2016,doi:10.1021/acs.jctc.8b00932,Huggins_2022,doi:10.7566/JPSJ.90.032001}. Variational quantum eigensolver (VQE) is a promising one among them. VQE was first developed by A. Peruzzo, et al. in 2014 \cite{peruzzo2014variational}. The underlying theory of VQE was explored in \cite{McClean_2016}. It utilizes the power of quantum computer to build a parameterized quantum circuit, and trains the parameters of the circuit with classical optimization approaches to minimize the energy of the represented quantum state for the studied Hamiltonian. The converged states represented by the quantum circuit is an approximation of the ground state of the studied system. VQE is a variational-type approach. The difference between VQE and classical variational approach (variational Monte Carlo \cite{becca_sorella_2017} for example) is that the most time-consuming part, like the calculation of the energy of a given quantum state, is performed on quantum computer.

VQE was tested on many few-body problems in physics and chemistry in the past few years (see \cite{TILLY20221, Cerezo_2021} for reviews).
Another area VQE could play an important role in is the study of strongly correlated quantum many-body systems. Exotic quantum states can emerge in these systems which enlarge our understanding of quantum phases and the phase transitions between them \cite{sachdev_2011}. However, the accurate determination of the ground state of general strongly correlated quantum many-body systems are challenging today even though a lot of many-body methods were developed \cite{PhysRevX.5.041041}. For example, the existence and the characterization of quantum spin liquid ground state on frustrated Heisenberg model is still under debate \cite{PhysRevLett.121.107202,PhysRevX.11.031034,doi:10.1126/science.1201080,PhysRevLett.109.067201,PhysRevX.7.031020,liao2017gapless}. Tremendous efforts have been devoted to the accurate determination of the ground state of slightly doped Hubbard model on square lattice \cite{doi:10.1126/science.aam7127,PhysRevX.10.031016,PhysRevB.102.041106,2023arXiv230308376X}, which is believed to be related to the high-Tc superconductivity in cuprates. There are already many attempts to apply VQE to the study of Heisenberg and Hubbard model on different lattice geometries \cite{PhysRevResearch.1.023025,kattemolle2022variational,PhysRevB.105.094409,cade2020strategies,PhysRevApplied.14.014059,fujii2022deep,Stanisic_2022}.

In this work, we investigate the performance of VQE on the antiferromagnetic Heisenberg model on kagome lattice with both nearest and next-nearest neighboring interactions. We simulate the VQE algorithm on classical computer without considering the gate noise. The antiferromagnetic Heisenberg model on kagome lattice was extensively studied previously because the ground state of this model is a possible quantum spin liquid due to the geometry frustration. We don't intend to solve this problem but use this difficult system as a test model to analyze the accuracy of VQE. From the exact diagonalization (ED) results of $J_1-J_2$ Heisenberg model on the kagome lattice, we find a level-crossing between the ground state and first excited state at $J_2/J_1 \approx 0.01$, indicating the existence of a first order phase transition near $J_2/J_1 \approx 0.01$ in the model. With VQE, we calculate the ground state energy for the parameter range of $-0.3 < J_2/J_1 < 0.3$ and compare them with the ED results. We find VQE energies agree well with the exact values in most region of parameters for the 18-site system we studied. However, near the phase transition point, VQE tends to converge to the excited states when the number of variational parameters is not large enough. We find this issue can be solved by either increasing the number of parameters in VQE ansatz or by initializing the parameters with converged values from nearby $J_2/J_1$ values. This issue can be encountered in the actual application of VQE on real quantum computer, so our work provides useful guidance for the future applications of VQE. 

The issue of converging to the exited states is in some sense different from the so-called barren plateau phenomenon \cite{mcclean2018barren} in VQE. Barren plateau means the gradient of the cost function or energy drops to zero rapidly with the increase of system size, which makes the optimization of the variational parameters harder and harder. What we found in this work is actually a local minimum problem.

The rest of the paper is organized as follows. In Sec.~\ref{model}, we introduce the model studied in this work. In Sec.\ref{exact}, we discuss the results from ED for the system with $18$, $24$, and $30$ sites. In Sec.~\ref{VQE}, we show the VQE results and discuss their accuracy. We also introduce different ways to solve the problem of converging to excited state in VQE near the first order phase transition point. In Sec.~\ref{con}, we conclude the work with a brief summary.

\section{Model}

\label{model}

The model studied in this work is the anti-ferromagnetic Heisenberg model on kagome lattice with both nearest ($J_1$) and next nearest ($J_2$) neighboring interactions with Hamiltonian 
\begin{equation}
\mathcal{H}=J_{1} \sum_{\langle i, j\rangle} \mathcal{S}^{(i)} \cdot \mathcal{S}^{(j)}+J_{2} \sum_{\langle\langle i, j\rangle\rangle} \mathcal{S}^{(i)} \cdot \mathcal{S}^{(j)}
\label{Hamiltonian}
\end{equation}
Both $J_1$ and $J_2$ are positive. We set $J_1$ as the energy unit in this work so there is only one free parameter $J_2 / J_1$ in the model.
\begin{figure}[t] 
\includegraphics[width=80mm]{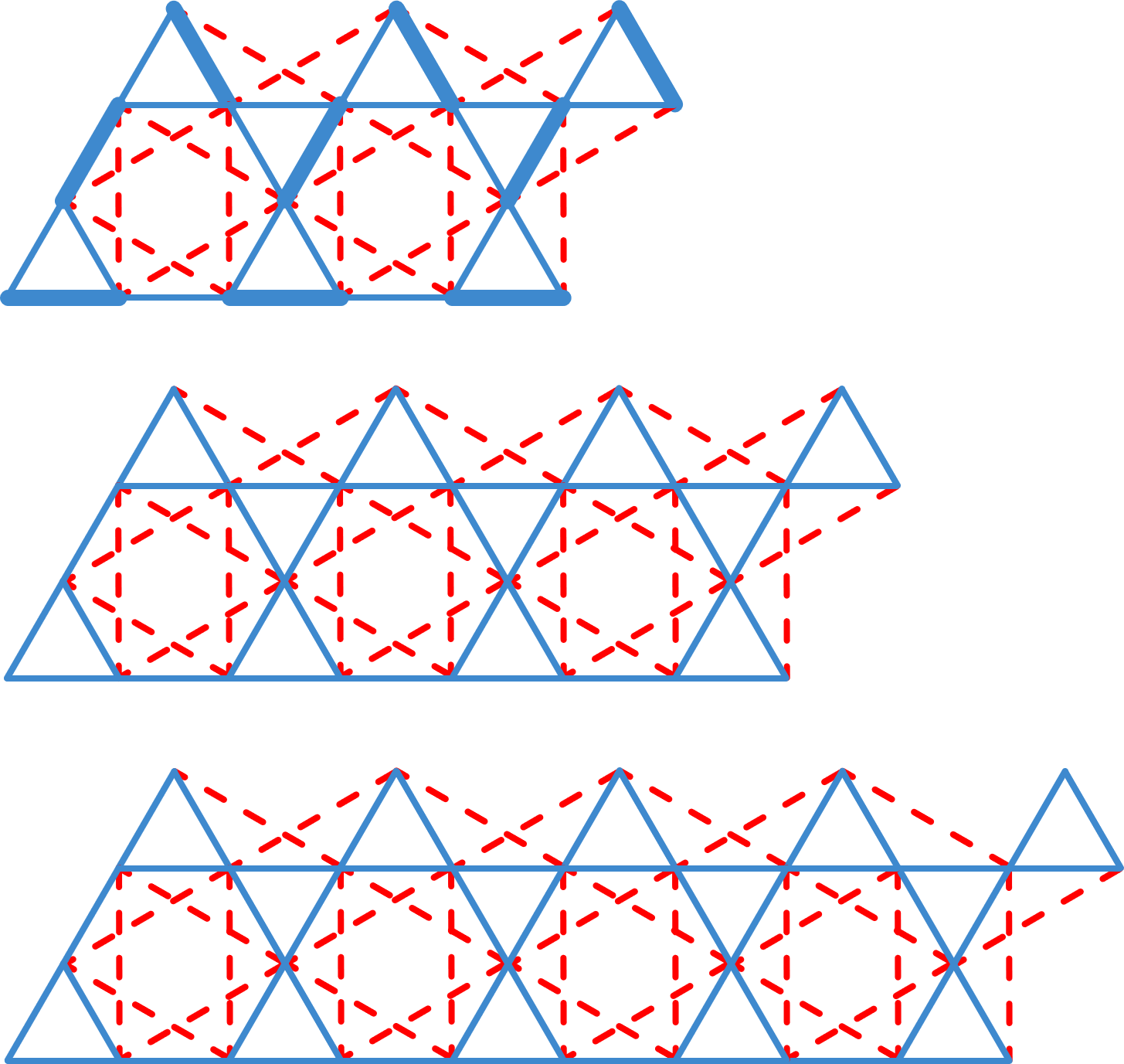}
\caption{Illustration of the kagome lattice with $18$, $24$, and $30$ sites studied in this work. Periodic boundary conditions are imposed. The solid blue lines represent the nearest neighboring $J_1$ terms and the red dashed lines represent the next nearest neighboring $J_2$ terms. In ED calculations, we study all the three systems. In VQE calculation, we only study the system with size $18$ in the top panel. The initial state in VQE is set as the product of the singlet states represented by the thick lines in the top panel.}
\label{kagome lattice}
\end{figure}

The anti-ferromagnetic Heisenberg model on kagome lattice was extensively studied before both with \cite{PhysRevB.91.020402,PhysRevB.91.104418,liao2017gapless} and without $J_2$ term \cite{doi:10.1126/science.1201080,PhysRevLett.109.067201,PhysRevX.7.031020,liao2017gapless}. Because the kagome geometry is highly frustrated, there are massive low-lying states near ground state, which makes it difficult to obtain the accurate ground state. The possible quantum spin liquid ground state in the model attracted a lot of attention, even though whether the ground state is a quantum spin liquid or not and the type of quantum spin liquid are still under debate \cite{doi:10.1126/science.1201080,PhysRevLett.109.067201,PhysRevX.7.031020,liao2017gapless}. We don't intend to solve this problem in this work but try to use this difficult model to test the performance of VQE. In this work, we study systems with $18$, $24$ and $30$ sites under periodic boundary conditions as illustrated in Fig.~\ref{kagome lattice}.

\begin{figure*}[t]
	\includegraphics[width=1.0\linewidth]{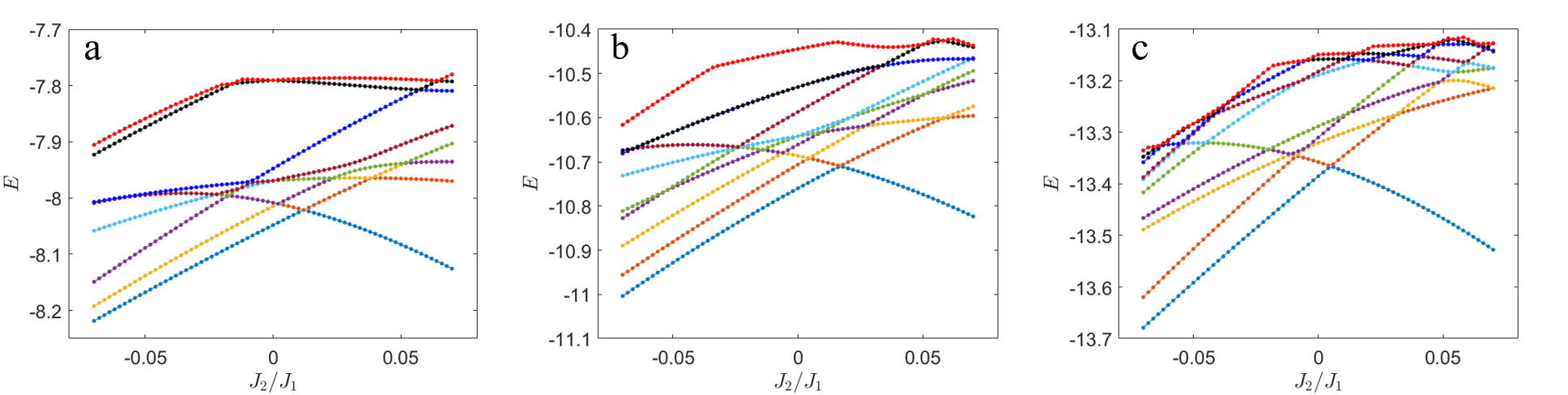}
	\caption{The lowest $10$ eigen-states in $S = 0$ sector for the systems with size $18$ (a), $24$ (b), and $30$ (c). We can find there are level crossings between the ground and first excited state for all three systems. The level-crossing points are $J_2/J_1 \approx 0.012, 0.016, 0.005$ for the system with size $18$, $24$, and $30$ respectively as plotted in Fig.~\ref{cross-sites}.}
	\label{Exact Diagonalization}
\end{figure*}

\section{Exact diagonalization results}
\label{exact}

In this section, we obtain the exact low energy eigen-states for system with sizes $18$, $24$, and $30$ (see Fig.~\ref{kagome lattice} for the details of the geometries) with ED method using the QuSpin package \cite{weinberg2019quspin}. In the ED calculation, we only use the conservation of $S_z^\text{tot}$. We list the lowest $10$ eigen-states in the $S = 0$ sector as shown in Fig.~\ref{Exact Diagonalization}. We notice that there are degeneracies in $S = 0$ sector.

\begin{figure}[b] 
\includegraphics[width=\linewidth]{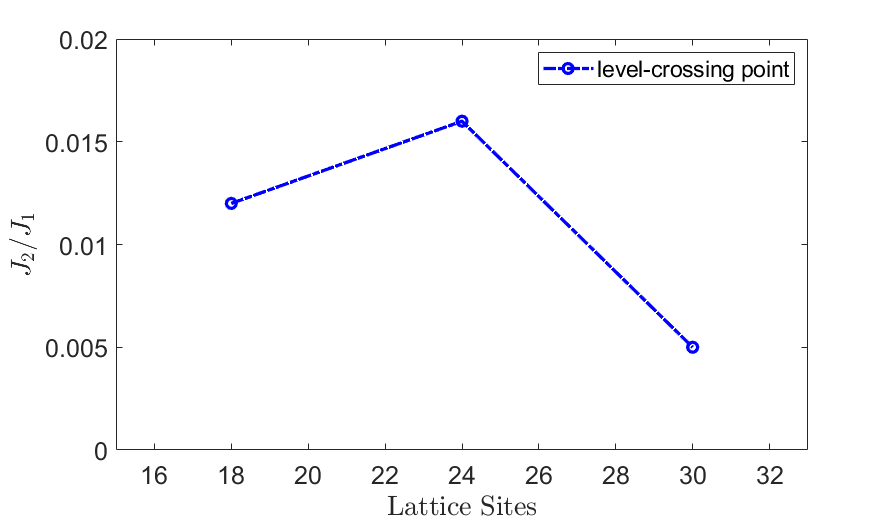}
\caption{The level crossing points for different system sizes as shown in Fig.~\ref{Exact Diagonalization}.}
\label{cross-sites}
\end{figure}

In Fig.~\ref{Exact Diagonalization}, we can find a level crossing between the ground state and first exited state for the three systems, which indicates a first order phase transition occurs in the system. The phase transition points are $J_2/J_1 \approx 0.012, 0.016$, and $0.005$ for systems with sizes $18$, $24$, and $30$ respectively as plotted in Fig.~\ref{cross-sites}. The determination of the precise location of the first order phase transition in the thermodynamic limit requires study on larger systems, which is beyond the goal of this work.

Previously, two phase transitions near $J_2 / J_1 = 0$ in the model were identified with different approaches \cite{PhysRevB.91.020402,PhysRevB.91.104418,liao2017gapless}, though there is discrepancy on the precise location of the phase transition points. Whether the first order phase transition found in this work is connected to the first or second phase transition observed in \cite{PhysRevB.91.020402,PhysRevB.91.104418,liao2017gapless} needs more investigation. Nevertheless, our ED results provide new insight about the nature of the phase transition in the model.

\section{VQE results}
\label{VQE}
We follow the setup in \cite{kattemolle2022variational} to perform the VQE calculation. We focus on the system with $18$-site as shown in the top panel of Fig.~\ref{kagome lattice}. We modify the HEISENBERGVQE code \cite{kattemolle2022variational} to study the Heisenberg model on kagome lattice with both $J_1$ and $J_2$ terms. We employ time-evolution-operator-like gate $\operatorname{HEIS}(\theta)\equiv \mathrm{e}^{-\mathrm{i} \theta / 4} \mathrm{e}^{-\mathrm{i} \theta \mathrm{S}^{(1)} \cdot \mathrm{S}^{(2)}}$ same as in \cite{kattemolle2022variational}. Each layer consists of the application of the gate $\operatorname{HEIS}(\theta)$ on every bond (including both the nearest and next nearest bonds) of the kagome lattice, which involves multiple gates in physical realizations \cite{kattemolle2022variational}. There are totally 72 gates for one layer of the quantum circuit. The initial state is chosen as the product of $9$ pairs of singlet states (the thick lines in the top panel of Fig.~\ref{kagome lattice}), so VQE state evolves in the sector of $S=0$ because the gate $\operatorname{HEIS}(\theta)$ conserves the SU(2) symmetry. For all the VQE calculations in this work, we initialize the parameter $\{\theta\}$ randomly if not otherwise specified. The BFGS method is then used to update the parameters $\{\theta\}$ till convergence. 

\begin{figure}[t]
\includegraphics[width=80mm]{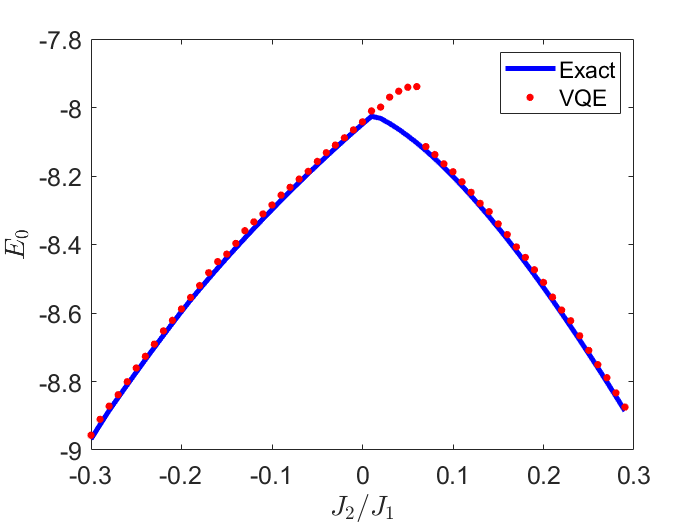}
\caption{The ground state energy for different $J_2/J_1$ values. The blue line are the ED results while the red dots are the VQE results. The VQE energies agree well with the exact values except for the region close to the level crossing point. See the main text for more discussion.}
\label{groundstate}
\end{figure}

In Fig.~\ref{groundstate}, the VQE energies for $-0.3 < J_2/J_1 < 0.3$ with $1200$ parameters is plotted as the red dots. The ED results are also plotted (the blue line) for comparison. It was found in \cite{kattemolle2022variational} that the VQE energy converges exponentially with the depth of the gate without $J_2$. In Fig.~\ref{groundstate}, the VQE energies for most $J_2$ values agree well with the exact results. However, there is significant deviation to the exact results for $J_2$ close to the first order phase transition point. By comparing the VQE energy in Fig.~\ref{groundstate} and the ED results in Fig.~\ref{Exact Diagonalization} (a), we find the trend of VQE energy for $J_2/J_1$close to the phase transition point is likely to follow one branch which crosses with another one at the phase transition point. We will have a detailed analysis of the VQE state for these $J_2/J_1$ values below. 

\subsection{Characterization of the VQE states}

To analyze the issue of large errors of VQE energy near the first order phase transition point, we calculate the components of exact low-lying eigen-states (from ED) in the state obtained by VQE for the system with $18$ sites. In Table \ref{Table 1}, we list the square of the overlap of the VQE states with the lowest $5$ eigen-states from ED. As can be found in Table \ref{Table 1}, the component of ground sate is almost zero for $0.02< J_2 < 0.06$ which means VQE with $1200$ parameters converges to the excited states instead of the ground state. For $J_2 / J_1 = 0.02$ and $0.03$, the most dominant component in the converged VQE state is the first excited state (with weight $0.99987$ and $0.9843$ respectively). For $J_2 / J_1 = 0.04$, $0.05$, and $0.06$, there are also considerable components of higher excited states in the converged VQE state. 

So the large error in VQE energy close to the phase transition shown in Fig.~\ref{groundstate} results from the convergence to excited states, while for the region far away from the phase transition point, the error comes from the tiny component of excited states in VQE (see the $J_2/J_1 = 0.01$ result in Table \ref{Table 1}). In Fig.~\ref{groundstate}, it is found that the VQE energy for $0.02< J_2 < 0.06$ is likely to follow one branch which crosses with another one at the first order phase transition point. This is consistent with the result that VQE converges to the first excited state in this region. 

%\C{So the large error in VQE energy on the right side close to the phase transition point as shown in Fig.~\ref{groundstate} results from the convergence to excited states, while for the region on the left side close to the phase transition point, the error comes from the tiny component of excited state in VQE (see the $J_2/J_1 = 0.01$ results in Table \ref{Table 1}). In Fig.~\ref{groundstate}, it is found that the VQE energy for $0.02< J_2 < 0.06$ is likely to follow one branch which crosses with another one at the first order phase transition point. This is consistent with the result that VQE converges to the first excited state in this region.}

\begin{table}[b]
\caption{The square of the overlap of the state obtained by VQE (with 1200 parameters) and the exact eigen-states.}
\begin{tabular}{|c|c|c|c|c|c|c|}
\hline $J_2/J_1$&0.01&0.02&0.03&0.04&0.05&0.06\\
\hline gs&0.83663&4.95e-05&1.66e-06&1.95e-05&3.08e-06&3.78e-06\\
\hline 1st&2.56e-06&0.99987&0.96430&0.65515&0.42972&0.66914\\
\hline 2nd&0.15609&7.63e-06&3.84e-04&6.39e-03&8.62e-04&0.26856\\
\hline 3rd&1.07e-03&3.37e-06&1.17e-03&0.01640&0.40520&0.03437\\
\hline 4th&8.08e-05&1.16e-06&0.02205&0.29664&0.06113&2.00e-05\\
\hline
\end{tabular}
\label{Table 1}
\end{table}

\subsection{Solutions to the issue of converging to excited states}
We propose two criteria for the cause of the problem of converging to excited states in VQE near the first order phase transition point. In the first criterion, with only $1200$ parameters, the variational space of VQE is limited and the ground state can't be reached no matter how well the parameters are optimized. If this is true, we can try to solve the issue by increasing the number of parameters. {Another criterion is that the randomly chosen initial parameters cause the optimization to be trapped to local minimums}. In this case, we can try to vary the initial values of the VQE parameters to solve this issue.

\subsubsection{VQE results with more parameters}

\begin{figure}[t] 
\includegraphics[width=80mm]{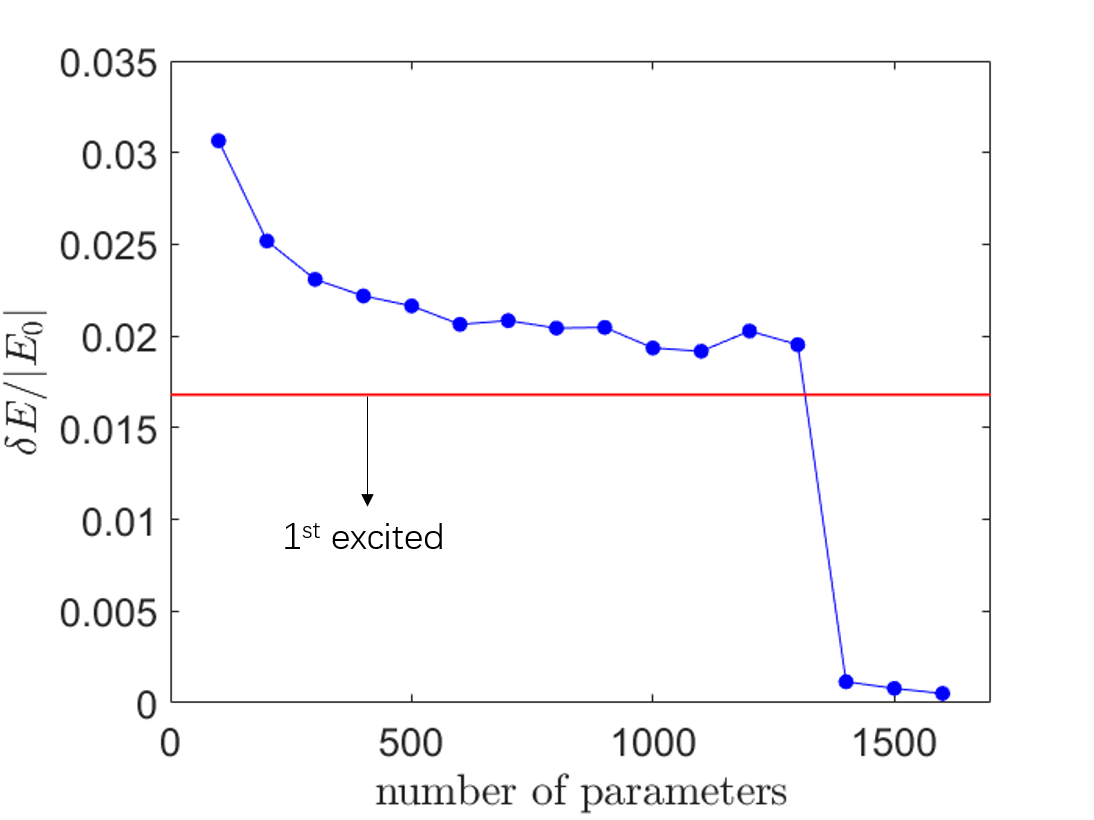}
\caption{The relative energy error $\delta E/|E_0|$ ($\delta E=E_0-E(\theta^*)$) of VQE for $J_2/J_1=0.06$ with different number of parameters. The blue dots represent the VQE results. The solid red line represents the result for the first excited state.}
\label{kagome_0.06}
\end{figure}

In Fig.~\ref{kagome_0.06} we show the relative error of VQE energy for the system with $18$ sites at $J_2 / J_1 = 0.06$. The red line marks the relative difference of the first excited state energy to the ground state energy. 
We can see that with the increase of the number of parameters, the energy first converges to an energy close to the first excited state and remains near this value for number of parameters less than $1300$. This is consistent with the results in Table~\ref{Table 1}. With the further increase of the number of parameters, the VQE energy finally converges to a value very close to the exact ground state energy. 

In Table~\ref{Table 2}, we list the evolution of VQE energy with the number of parameters for $J_2/J_1 = 0.01\sim0.06$. We find that with the increase of the number of parameters, VQE is able to get out of the local minimum of the first excited state and the energy indeed converges to the ground state except for the $J_2 / J_1 = 0.02$ case, which remains in the excited state for the maximum number of parameters we can reach. We also find that the required number of parameters to be able to converge to the ground state increases when $J_2 / J_1$ gets closer to the phase transition point, which is reasonable because the excitation gap closes gradually in this process. 
\begin{table}[b]
\caption{The energies obtained with VQE with different number of parameters for $J_2/J_1 = 0.01 \sim 0.06$. The exact energies for the ground and first excited state are also listed for comparison. For almost all the cases, the VQE state converges to the ground state with the increase of number of parameters. The $J_2/J_1 = 0.02$ case remains in the excited state for the largest number of parameters we can reach.}
\begin{tabular}{|c|c|c|c|c|c|c|}
\hline $J_2/J_1$ &0.01&0.02&0.03&0.04&0.05&0.06\\
\hline 1200&-8.01005&-7.99874&-7.96990&-7.95254&-7.94093&-7.94513\\
\hline 1800&-8.02381&-7.99982&-7.97851&-7.95955&-7.95314&-8.09919\\
\hline 3600&-8.02513&-8.00290&-7.98163&-8.06289&-8.08241&-8.10310\\
\hline 4800&/&-8.00340&-8.04656&/&/&/\\
\hline $E_0^{ED}$&-8.02573&-8.03170&-8.04671&-8.06374&-8.08266&-8.10333\\
\hline $E_1^{ED}$&-8.01885&-8.00368&-7.98217&-7.96443&-7.96537&-7.96719\\
\hline
\end{tabular}
\label{Table 2}
\end{table}

\subsubsection{VQE results with different initial parameters}

To test the second criterion that the converging to excited state problem near the phase transition point is caused by the randomly initialization of the VQE parameters, we use the $1200$ converged VQE parameters for $J_2/J_1 = 0.07$ to initialize the VQE parameters for $0.02\leq J_2/J_1\leq 0.06$. From Fig.~\ref{groundstate}, we can see that the VQE energy for $J_2/J_1 = 0.07$ is converged to the ground state. The square of the overlap of the VQE state from this new initial parameters with the exact ground state from ED are shown in Table~\ref{Table 3}.

In Table~\ref{Table 3}, we find the VQE states all converge to the exact ground state for $J_2/J_1=0.02\sim0.06$. In these calculations, the number of parameters is $1200$, same as the parameter number used to obtain Fig.~\ref{groundstate}. These results show that the converging to the excited state problem indeed results from the randomly initialization of the VQE parameters. Choosing suitable initial parameters for VQE was also extensively discussed in the literature \cite{TILLY20221, Cerezo_2021, grimsley2022adapt}.

\begin{table}[t]
\caption{The square of the overlap of the state obtained by VQE (with 1200 parameters) and the exact ground state. The parameters in VQE are initialized with the converged values for $J_2 /J_1 = 0.07$.}
\begin{tabular}{|c|c|c|c|c|c|}
\hline $J_2/J_1$&0.02&0.03&0.04&0.05&0.06\\
\hline gs&0.99591&0.99397&0.99247&0.99159&0.99104\\
\hline
\end{tabular}
\label{Table 3}
\end{table}

The above discussion indicates that the first criterion that the ground state can't be reached due to limited variational space is not correct. But in practical calculation, the converging to excited state problem for the studied $18$-site system can be solved by either increasing the number of parameters or trying different initialization of the parameters. Our discussion indicates that care should be taken when interpreting the VQE results because in practical application of VQE to solve unknown problems, it is difficult to judge whether the number of parameters is large enough or not.

\section{Conclusion}
\label{con}

We test the performance of VQE by studying the $J_1$-$J_2$ Heisenberg model on the kagome lattice. We simulate the algorithm in classical computer without considering the effect of gate noise. The VQE energies agree well with the exact results in most parameter regions. But we find VQE tends to converge to the excited states near the first order phase transition point when the number of variational parameters is not large enough. This problem can be solved by either increasing the number of variational parameters in VQE or by initializing the parameters using the converged values away from the phase transition point. Our work indicates that when performing practical VQE calculation on real quantum computer, we need to try different initial values for the variational parameters and check the convergence of the results with the number of parameters to avoid the problem of converging to excited states.

\begin{acknowledgments}
MQ acknowledges the support from the National Natural Science Foundation of China (Grant No. 12274290) and the sponsorship from Yangyang Development Fund.
\end{acknowledgments}

\bibliography{vqe.bib}

\appendix

\end{document}